\documentstyle[aps,prl,psfig,twocolumn]{revtex}

\begin{document}

\title{On the microcanonical solution of a system of fully coupled particles}

\author{Micka{\"e}l Antoni \thanks{E-mail: antoni@ms531u07.u-3mrs.fr}$^{1}$,   
Haye Hinrichsen\thanks{E-mail: hinrichs@comphys.uni-duisburg.de}$^2$  
and Stefano Ruffo\thanks{E-mail: ruffo@avanzi.de.unifi.it}$^{3}$}

\address{$^1$ Laboratoire de Thermodynamique, Universit\'e d'Aix-Marseille III,  
Centre St. J\'er\^ome, Service 531, F-13397 Marseille, France}  
\address{$^2$ Theoretische Physik, Fachbereich 10, Gerhard-Mercator Universit\"at,  
47048 Duisburg, Germany}  
\address{$^3$ Dipartimento di Energetica ``S. Stecco'',   
Universit\`a di Firenze, INFM and INFN  
via S. Marta 3, I-50139 Firenze, Italy}   
\date{\today}  
  
\draft  
\maketitle  
\begin{abstract}  
We study the Hamiltonian Mean Field (HMF) model, a system of $N$  
fully coupled particles, in the microcanonical ensemble. We use  
the previously obtained free energy in the canonical ensemble to  
derive entropy as a function of energy, using Legendre transform  
techniques. The temperature-energy relation is found to coincide  
with the one obtained in the canonical ensemble and includes a  
{\it metastable} branch which represents spatially homogeneous  
states below the critical energy. ``Water bag" states, with removed  
tails momentum distribution, lying on this branch, are shown to  
relax to equilibrium on a time which diverges linearly with $N$  
in an energy region just below the phase transition.  
\end{abstract}  
\pacs{PACS numbers:05.40+j,05.70Fh}

\section{Introduction}  
In a pioneering work Fermi, Pasta and Ulam~\cite{fer1} showed   
that in a Hamiltonian system with {\it short-range} interactions   
the relaxation time to thermodynamical equilibrium can be extremely   
long if the energy is small enough. This striking feature triggered   
an intense research activity and, nowadays, the relation between   
macroscopic thermodynamical properties and microscopic   
dynamics is actively investigated in a large variety   
of physical systems~\cite{chaos}.  
The common belief that microcanonical time averages   
coincide with equilibrium ensemble averages after a possibly  
quite long transient is confirmed by several numerical  
experiments (see Refs.~\cite{ruf0} for reviews). Therefore, although  
the rigorous proof of {\it ergodicity} is restricted to a few rather  
abstract dynamical systems (e.g. the Sinai billiard), it is  
nevertheless believed, and confirmed numerically~\cite{Esc,Giard}, that   
time averages of thermodynamic functions (temperature,   
internal energy, specific heat)   
converge to their equilibrium values, at least in the regime of strong chaos.  
  
The case of {\it long-range} forces is quite less explored from this point  
of view. The study of relaxation to thermal equilibrium is here made  
more complex by the consequences of the {\it inequivalence of statistical  
ensembles}. Indeed, it is well known that the impossibility to   
isolate a small sub-system from a large thermal bath, due to the  
presence of long-range forces, prevents the derivation of  
{\it canonical} from {\it microcanonical} ensemble.  
 
Exactly solvable toy models~\cite{Thirring} with weak long-range forces  
explicitly display ensemble inequivalence and the presence of  
{\it negative specific heat} in the microcanonical ensemble.  
Similar phenomena were also obtained for gravitational models~\cite{Lynden-Bell}.  
These toy models show a {\it first order phase-transition} in the   
canonical ensemble, which corresponds to the presence of a concave region  
of the entropy-energy curve in the microcanonical ensemble.  
This feature is present also in microcanonical molecular dynamics  
simulations of {\it short-range} models without hard-core~\cite{Compagner},  
where one observes, in the phase-transition region, a high-density  
{\it clustered} phase coexisting with a low-density {\it gaseous} phase.  
More recently, microcanonical Monte-Carlo simulations have shown  
similar properties for the diluted three-state Potts model~\cite{Gross}.  
  
All the previous remarks concern {\it equilibrium} properties, but it is  
known that long-range interacting systems show also extremely slow  
relaxation to equilibrium. A remarkable example is the one-dimensional  
self-gravitating potential, where the study of long transients and  
quasi-equilibria has been undertaken by many groups~\cite{Miller}.  
More recently, a scaling law for the relaxation to equilibrium has  
been proposed for gravitational systems based on diffusion in   
phase-space~\cite{Pomeau}.  
    
The Hamiltonian Mean Field (HMF) model~\cite{Ruffo} has been  
introduced with the aim of studying {\it clustering} phenomena  
in $N$-body systems in one dimension. The original motivation was to  
consider a truncated Fourier series development of the one-dimensional  
gravitational and charged sheet models. Its canonical ensemble exact   
solution predicts a {\it second-order} phase transition from a   
clustered phase to a gaseous one.  
HMF is an {\it infinite range} interacting system and one would  
naturally expect unusual thermodynamical behaviour. However, microcanonical simulations 
of HMF show that ``carefully" prepared initial states lead to a 
good agreement with canonical analytical results for the energy-temperature  
relation~\cite{Latora}. 
On the other hand, ``water bag" initial states, {\it {ie.}} states with 
momentum distribution on a finite support, are {\it metastable}  and  
show an extremely slow relaxation to equilibrium near the  
critical energy~\cite{Latora}.  
  
In this paper, we derive analytically the entropy-energy relation   
of the HMF model in the microcanonical ensemble. We further study  
the properties of  
metastable states corresponding to the low energy extension of the high energy   
gaseous phase. The relaxation time of these states to canonical equilibrium   
is also investigated, showing that it increases linearly with   
the number of particles $N$ near the critical energy.

\section{The Model}  
The Hamiltonian Mean Field model describes the dynamics of $N$ classical   
and identical particles confined to move on the unit circle~\cite{Ruffo}.   
Its Hamiltonian writes  
\begin{equation}  
H= \sum_{i=1}^N \frac {p_i^2} 2   
+ \frac 1 {2N} \sum_{i,j=1}^N \Bigl(1-\cos(\theta_i-\theta_j)\Bigr)=K+V,  
\label{ham}  
\end{equation}  
where $(\theta_i,p_i)$ are canonically conjugated coordinate and momentum 
variables with  $\theta_i \in [0,2\pi[$. $K$ is the kinetic energy and the potential  
energy $V$ corresponds to the first harmonic of the one-dimensional self-gravitating  
potential   
$V=\sum_{i,j}|\theta_i-\theta_j|$ (up to scaling constants). Total momentum   
$\sum_i p_i$ is also conserved and always set to zero without loss of generality.  
  
In the {\it canonical} ensemble, model (\ref{ham}) undergoes a   
second-order phase transition at the inverse temperature   
value $\beta_c=T_c^{-1}=2$~\cite{Ruffo}, corresponding to a critical  
specific energy $U_c=E_c/N=3/4$. The free energy   
\begin{equation}  
F(\beta)= - \frac{1}{\beta} \lim_{N \to \infty} \Biggl[\frac{1}{N} \ln Z (\beta,N)\Biggr]~,  
\label{free}  
\end{equation}  
where   
\begin{equation}  
Z(\beta,N)=\int \prod_i dp_i d \theta_i \exp (-\beta H)~,  
\end{equation}  
has been obtained using saddle-point techniques~\cite{Ruffo}, giving  
\begin{equation}  
-\beta F= \frac{1}{2} \ln\biggl[ \frac{2 \pi}{\beta} \biggr]-\frac{\beta}{2}  
+\max_x \biggl[X(x,\beta)\biggr]~,  
\label{solution}  
\end{equation}  
where   
\begin{equation}  
X(x,\beta)=\ln (2 \pi I_0(x)) -\frac{x^2}{2 \beta}~,  
\label{var}  
\end{equation}  
$I_0$ being the zero order modified Bessel function.  
The maximization of $X$ is performed by solving the consistency equation  
\begin{equation}  
\frac{x}{\beta}=\frac{I_1(x)}{I_0(x)},  
\label{consistency}  
\end{equation}  
whose unique solution at small $\beta$ is $x=0$, while at large $\beta$,  
{\it ie.} low temperature, a pair of symmetric non vanishing solutions is present.  
The bifurcation (of pitchfork type) occurs at $\beta_c=2$ and  
at this temperature the value of the free energy F has a discontinuity  
in the second order derivative, which is the 
signature of a second-order phase transition.  
This is also confirmed by the calculation of the order parameter  
\begin{equation}  
{\bf M} = \frac 1 N   
\Biggl( \sum_{i=1}^N \cos\theta_i,\sum_{i=1}^N \sin\theta_i\Biggr)  
= M \Bigl( \cos \Phi, \sin \Phi \Bigr),  
\label{meanfield}  
\end{equation}  
which vanishes continuously at $\beta_c$, remaining zero in the   
whole high temperature phase.   
At low temperatures, $M$ measures the degree of clustering   
of the particles.  
  
The equations of motion derived from model (\ref{ham}), written in terms of   
${\bf M}$, show that each particle has a pendulum like motion  
\begin{equation}  
\ddot{\theta_i} = - M \sin (\theta_i - \Phi)~,  
\label{ham2}  
\end{equation}  
which makes explicit the self-consistent nature of the dynamics, since  
each particle moves in the M-field, which is itself determined by  
the position $\theta_i$ of all the particles.  
This has led~\cite{Ruffo} to the interpretation of the phase transition  
as a dynamical process of particle evaporation from the cluster. 
Moreover, this approach follows the idea that the study of self-consistent  
$N$-body dynamics is simplified by an ``effective" reduction of the number of degrees  
of freedom~\cite{Tennyson}.  
  
\section{Entropy - Energy relation.}  
The free energy $F(\beta)$ in the canonical ensemble is readily computed  
once one knows the dependence of entropy $S$ on 
specific energy $U=E/N$ in the microcanonical ensemble. One begins from the relation  
\begin{equation}  
Z(\beta,N)= \int_{0}^{\infty} \omega(E,N) e^{-\beta E}dE,  
\label{part}  
\end{equation}  
where  
\begin{equation}  
\omega(E,N)= \int \prod_i dp_i d \theta_i \delta (E-H)  
\end{equation}  
is the microcanonical phase-space density at energy $E$. The lower  
limit of the integral in (\ref{part}) is given by the energy of  
the ground state, which vanishes in our model (in this case all of the  
particles are at the same position with zero momentum).  
After rescaling by $N$ and exponentiating $\omega$ Eq. (\ref{part}) becomes  
\begin{equation}  
Z(\beta,N) = N \int dU \exp \bigl( N(-\beta U+ \frac{1}{N} \ln (\omega(E,N)) \bigr)~.  
\end{equation}  
This integral is then solved by the saddle-point technique in the  
$N \to \infty$ limit, which implies taking the maximum of the  
argument of the exponential. Using the definition of entropy  
\begin{equation}  
S(U)= \lim_{N \to \infty} \Biggl[\frac{1}{N} \ln \omega(UN,N) \Biggr] 
\label{ent}  
\end{equation}  
and recalling the definition of free energy (\ref{free}), one  
finally gets  
\begin{equation}  
- \beta F(\beta)=\max_U [-\beta U + S(U)]~.  
\label{legendre}   
\end{equation}  
Entropy $S$ and free energy $F$ are thus related by a Legendre  
transform. {\it Assuming} now that $S$ is concave ({\it i.e.} 
downward bended)\footnote{We thank  
an anonymous referee for this remark.}, one can invert (\ref{legendre}) and obtain  
\begin{equation}  
S(U)=\min_{\beta > 0}[\beta (U - F(\beta))].  
\label{inv_legendre}  
\end{equation}  
This assumption is not innocent and the invertion (\ref{inv_legendre}) can be safely   
used only for second order phase-transitions, as it is for our  
model. On the contrary, it is well known that for first-order   
phase transitions entropy has a concave region~\cite{Thirring,Gross},   
which is responsible for the negative specific heat in the   
microcanonical ensemble. A generalization of model (\ref{ham})  
to particle motion on the two-dimensional torus~\cite{Antoni}  
does have a first-order phase transition, making the invertion  
impossible in this case.  
  
\begin{figure} [h]  
\centerline{\psfig{figure=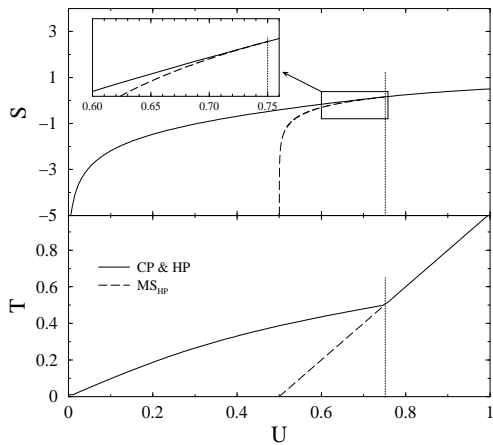,angle=270,width=7.0cm,height=6.0cm}}
\caption{The entropy $S$ as a function of $U$   
(upper graph) and the corresponding energy temperature relation (lower graph).  
The equilibrium values of $S$ and $T$ vs. $U$ is plotted as a solid line,  
both above (HP) and below (CP) the critical energy $U_c=3/4$. These same  
quantities, evaluated for the metastable homogeneous phase $MS_{HP}$, are   
represented by dashed lines. The inset in the upper graph zooms in the   
energy range where the phase transition occurs. The vertical dotted  
line indicates the critical energy $U_c$.}  
\label{fig1}  
\end{figure}

The calculation of entropy in the microcanonical ensemble is then  
reduced to the min-max procedure  
\begin{equation}   
S(U)= \min_{\beta > 0} \max_x [\beta U + \frac{1}{2} \ln \bigl( \frac{2 \pi}{\beta} \bigr)  
-\frac{\beta}{2} + X(x,\beta) ]~,  
\label{solution2}  
\end{equation}  
with $X$ given by (\ref{var}). For $\beta > \beta_c=2$ the {\it max} is  
at $x=0$ and the {\it min} at $\beta = 1/(2U-1)$, which gives the   
entropy function  
\begin{equation}  
S(U)= \frac{1}{2} \ln (2U -1) + \mbox{const.}~.  
\label{HP}  
\end{equation}  
This is plotted in fig.~\ref{fig1} with full line above $U=U_c=3/4$ and  
with dashed line for $1/2 < U < U_c$ (below $U=1/2$ the   
temperature $T=1/\beta$ becomes negative and the solution meaningless).  
This solution corresponds to the high temperature Homogeneous Phase ($HP$)   
with $M=0$ for $U>U_c$ and to the {\it MetaStable} (the meaning is  
clarified below) Homogeneous Phase $MS_{HP}$ below $U_c$. Both  
phases are present also in the canonical solution and   
while in this latter context the $MS_{HP}$ corresponds to a local, but not  
global, minimum of the free energy, in the microcanonical context the  
entropy of the $MS_{HP}$ is not maximal.   
Indeed, the solution which maximizes the entropy bifurcates continuously  
from this one at $U_c$. Below this energy the equilibrium solution is related to the two symmetric  
solutions with $x \neq 0$ of the consistency equation (\ref{consistency}).  
They give the new maximum value of $X$ in (\ref{solution2}) and  
the numerical  
search of the minimum in $\beta$, which is unique, allows to compute  
the stable branch of $S(U)$ below $U_c$. The latter is plotted in fig.~\ref{fig1}  
with the continuous line and labelled $CP$, because it corresponds to  
the Clustered Phase, which is characterized by a non vanishing value of $M$.   
The temperature $T$ can also be derived from the standard relation  
$\partial S/\partial U = T^{-1}$ and is reported in fig.~\ref{fig1}.    
The stable solution $S(U)$ (full line) is continuous with its  
derivative. The discontinuity is in the second derivative, which gives  
a jump in the specific heat $C_V = (\partial T/\partial U)^{-1}$.  
  
{\it To summarize, our results suggest that the microcanonical  
solution is fully consistent with the one   
in the canonical ensemble, and we can therefore conclude that the   
two ensembles lead to equivalent results for model (\ref{ham}).}

\section{Metastability}  
  
However, the dynamics reveals a quite peculiar behavior of the  
microcanonical {\it metastable} branch.  
We have performed numerical simulations in microcanonical  
fixed-energy conditions by integrating the equations   
of motion (\ref{ham2}), starting from appropriately chosen initial states.  
We have used a fourth order symplectic algorithm~\cite{Mac}  
with a time step $0.1$, allowing relative energy conservation   
as good as $10^{-5}$. We have checked the energy-temperature curve,  
definining temperature as twice the time average of the kinetic  
energy: $T=2<K>$.  
The equilibrium energy-temperature solution is reached from generic initial  
states over most part of the energy range. However, for   
$1/2 < U < U_c$ relaxation to equilibrium becomes very slow.    
We have started the runs in this range with the particles   
homogeneously distributed on the circle ($M=0$) and with a ``water bag'' momentum   
distribution: the momenta are uniformly distributed in an interval around zero  
such that the variance of the distribution coincides initially with the  
temperature of the {\it metastable} branch $T=2U-1$.  
   
\begin{figure} [h]  
\centerline{\psfig{figure=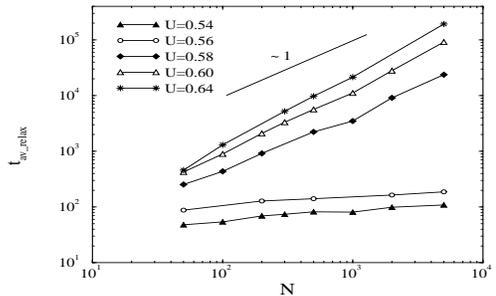,angle=270,width=7.0cm,height=4.0cm}}
\caption{Average relaxation time $t_{av\_relax}$ for an initial $MS_{HP}$   
configuration as a function of $N$ for several values of $U$   
with $\sigma=0.9$. The upper solid line indicates the   
slope of the linear $N$ dependence for comparison.}   
\label{fig2}  
\end{figure}

The relaxation of the initial ``water bag'' to the equilibrium state proceeds   
through the slow development of Maxwellian tails in the velocity distribution.   
In the literature one can already find indications that the relaxation time  
to equilibrium grows linearly with $N$~\cite{Latora} (the authors used  
as a criterion the convergence to the Boltzmann entropy of the Maxwellian).   
We have decided to consider directly the time evolution of the  
temperature of the $MS_{HP}$   
states. The initial temperature lies below the equilibrium one (see dashed  
line in fig. \ref{fig1}) and relaxes to a fixed threshold temperature $\sigma T_{eq}$   
after a time $t_{relax}$, where $\sigma < 1$ is the threshold amplitude  
and $T_{eq}$ is the microcanonical equilibrium temperature represented by   
the full line in fig.~\ref{fig1}. Considering different initial states, we  
have observed that the fluctuations of $t_{relax}$  
are Poissonian and that a well defined average exists.  
This average relaxation time $t_{av\_relax}$   
is plotted in fig.~\ref{fig2} for $\sigma=0.9$ as a function of $N$ for different  
values of $U$. For $U < 0.56$, $t_{av\_relax}$ is almost independent   
of $N$ whereas for larger values of $U$, $t_{av\_relax}$ increases  
approximately as $N$.  
Nothing substantially changes using different thresholds by changing the  
value of $\sigma$.   
We are therefore able to confirm, for a different observable and a better statistics  
(we have considered up to 3600 initial conditions for the points in fig.~\ref{fig2}),  
the results of Ref.~\cite{Latora} on the linear divergence of the relaxation 
time with $N$, at least in a restricted energy interval just below $U_c$.  
The existence of such divergence implies that performing the $N \to \infty$ limit  
before the $t \to \infty$ limit, the system will stay indefinitely out of equilibrium  
and the momentum distribution will never develop Maxwellian tails.  
This has attracted the attention of Tsallis~\cite{Tsallis}, who argues that  
his entropy definition should be appropriate to describe the thermodynamical  
properties of such {\it metastable} states in the thermodynamic limit (in  
particular it would possibly describe the power law tails of the momentum  
distribution).  
It should however be remarked that what we have tested here is not the divergence   
of the {\it lifetime} of such states, but that of the relaxation time. However,   
recent numerical simulations performed  
by Rapisarda~\cite{Rapisarda} seem to indicate also the divergence of the  
lifetime with $N$, encouraging Tsallis' interpretation.  
Furthermore, it is possible to prepare states on the metastable  
branch which have initially a Maxwellian momentum distribution and those  
states are observed to relax to equilibrium pretty fast. It seems that  
the ``water bag" condition, which corresponds to an initial truncation of  
the tails, is a necessary one for obtaining a divergence of the relaxation  
time.  
  
\section{Conclusions and perspectives}  
  
Let us summarize what we have found. The {\it microcanonical} solution of the   
Hamiltonian Mean Field model is here obtained with Legendre transform techniques,   
which avoid state counting.   
The entropy is computed using a min-max procedure and the energy-temperature   
relation is derived analytically.  
The equilibrium states are shown to coincide with the {\it canonical} ones   
in the thermodynamic limit. This result is not obvious   
since, due to the long-range interparticle coupling, ensemble equivalence  
might have been violated. Besides the equilibrium solutions, other states   
are found, corresponding to the {\it metastable} homogeneous configurations  
lying below the critical energy. The study of their  
relaxation time to equilibrium exihibits a divergence proportional to  
the number of particles, meaning that in the thermodynamic limit the trajectory  
in the phase-space remains indefinitely trapped in a non-equilibrium  
region (something similar to so-called ``mechanical" initial states 
for the Fermi-Pasta-Ulam model in Ref.~\cite{deLuca}).  
Further investigations are necessary to understand the relevance of  
the metastable states for other more realistic long-range  
forces, beginning with the very interesting extension of the HMF proposed by   
Tsallis and co-workers~\cite{Ante1,Ante2}, whose canonical solution  
has been recently obtained in Ref.~\cite{Giansanti} . These results could be   
relevant for the physics of self-gravitating systems and  
for nuclear dynamics, where long-range forces play a distinctive role.

\vskip 2 truecm  
{\bf Acknowledgements}

We thank  V. Latora, A. Rapisarda and A. Torcini for fruitful discussions  
and ongoing collaboration on this subject.  
M. A. acknowledges D. Noack for important logistics support. M.A. and S.R. thank the Institute  
for Scientific Interchange, Torino (Italy) and the Laboratoire de Physique of  
the ENS-Lyon (France) for hospitality and financial support.  
The Max-Planck Institute for Physics of Complex Systems in Dresden (Germany)  
is also acknowledged for allowing us to use its powerful computer center.

%%%%%%%%%%%%%%%%%%%%%  
%%%%%%%%%%%%%%%%%%%%%  
  
%%%%%%%%%%%%%%%%%%%%%%%%%%  
%%%%%%%%%%%%%%%%%%%%%%%%%%   

\end{document}